# A tool to support the investigation and visualization of cyber and/or physical incidents

Inês Macedo [0000-0002-0837-2410], Sinan Wanous [0000-0002-9711-4850], Nuno Oliveira [0000-0002-5030-7751], Orlando Sousa [0000-0003-0779-3480], Isabel Praça [0000-0002-2519-9859]

School of Engineering from the Polytechnic of Porto (ISEP)
Research Group on Intelligent Engineering and Computing for Advanced Innovation and Development (GECAD)
{incar, sinai, nunal, oms, icp}@isep.pt

**Abstract.** Investigating efficiently the data collected from a system's activity can help to detect malicious attempts and better understand the context behind past incident occurrences. Nowadays, several solutions can be used to monitor system activities to detect probable abnormalities and malfunctions. However, most of these systems overwhelm their users with vast amounts of information, making it harder for them to perceive incident occurrences and their context. Our approach combines a dynamic and intuitive user interface with Machine Learning forecasts to provide an intelligent investigation tool that facilitates the security operator's work. Our system can also act as an enhanced and fully automated decision support mechanism that provides suggestions about possible incident occurrences.

## 1    Introduction

Security investigation aims to explore the cause of an attack and how much it threatened the security of the targeted property. In case the security of a system is compromised, investigating over the data collected during the monitoring phase can bring important insights, both to improve detection and prevention, and to support mitigation and remediation strategies. Intrusion detection sensors can collect additional evidence, in the form of logs, related to system and network accesses, email services, file analysis, etc. In complex infrastructures, the number of systems can be huge, and thus the amount of data collected. The proper data visualization can be crucial to support the Security Operators in analyzing all the evidence collected. If we consider domains where both cyber and physical security monitoring solutions are used, not only the data volume is bigger, also the analyses of both domains are crucial to understand how complex attacks can be triggered, benefiting from the combination of cyber and physical threats.

In this work, we describe an investigation tool, that is based on the time series visualization of cyber and physical attacks, and the way a machine learning and a visualization module are interconnected to support the detection and correlation of incidents.

The paper is divided into multiple sections that can be detailed as follows. Section 2 provides an overview of related security frameworks, incident investigation research, and commercial security systems. Section 3 briefly describes the security ecosystem and our solution's internal architecture. In Section 4, our system's visualization component is illustrated with more detail, specifying the main functionalities as well as the most relevant dashboards that can be accessed. Finally, Section 5 provides a conclusion about how our approach facilitates the SOC Operator's work and enhances the overall security process.

## 2    Related Work

Authors of Thakur, K., et al. [1] describe crucial issues in cybersecurity domains. They focus on the issues of cyber security threats and summarizes the existing security models and research directions in each field. Moreover, a basic network infrastructure for applications and systems vulnerability analysis is proposed in [2]. It aims to start cybersecurity investigation in a real hardware manner. The goal of this infrastructure is to ensure secure systems, planning and operation, response, and support. However, in their work, Ussath, M., et al. [3] depict a concept for a security investigation framework to improve the efficiency of investigations of multi-stage Advanced Persistent Threat (APT). The proposed concept utilizes automatic and semi-automatic functions and requires core components such as parsing, visualization, meta information, and correlation components. Also, it collects relevant information from different sources, such as log files, results of forensic investigations, and malware analysis reports.

Another study explores the complex investigation of incidents in cloud-based environments [4], in which the automated monitoring tools may implicate rather long lists of virtual machines and containers. The authors propose a visualization approach that aims at reducing the number of VMs and containers awaiting forensic investigation. They conclude that, using the visualization tools, individuals were able to detect malware 70% of the time. Also, future combinations of visualizations with intelligent forensics might provide better results. Rondeau, C., et al. [5] try to improve pre-attack security and post-attack forensic methods in the context of WirelessHART - IIoT (Wireless Highway Addressable Remote Transducer in support of Industrial Internet of Things). They investigate activity aimed at applying Time Domain Distinct Native Attribute (TD-DNA) fingerprinting and improving feature selection to increase computational efficiency and the potential for near-real time operational application. A mini-track overview reports state-of-the-art in the emerging area of cybersecurity investigations and digital forensics [6]. According to the authors, there are still pending challenges in this area. Challenges include the identification of solutions to handle complex investigations, especially those that involve technology such as smart cities, cyber-physical systems, and Internet of Things (IoT) environments.

According to Kebande, V., et al. [7] there is a lack of holistic and standardized approaches addressing the digital forensic investigations of IoT systems. In their paper, authors developed and validated a Digital Forensic Readiness (DFR) framework to be used to implement a proactive forensic process in an organization. The presented framework is compatible with the ISO/IEC 27043:2015 standard and considers organizational complexities and policy developments. On the other hand, a report by Horsman, G. [8] takes a step forward in the field-wide sharing of knowledge in digital forensics. He proposes a framework designed to set out the required elements for sharing reliable digital forensic knowledge, called the Capsule of Digital Evidence (CODE).

Due to a study from 2015 by Sandia National Laboratories, USA [9], stakeholders of physical protection systems lack essential knowledge related to cyber-enabled physical attack scenarios. In this study, Sandia R&D team worked to support experiments of cyber-attacks against Physical Protection Systems (PPS) components and subsystems. Their research focuses on developing a reliable capability to investigate cyber-enabled physical attack scenarios: identifying and understanding credible cyber threats to PPS.

Furthermore, an emerging sub-discipline of digital forensics was described [10]. This field covers Financial Technologies (or Fintech), and recognize investigations related to financial transactions and payment activities. In another context, there is a rise in cybercrimes related to automated programs, in which, existing tools and research are not capable enough to advance bot crime investigations [11]. In their paper, Rahman, R. U., el al. [11] propose a four-phase web forensic framework to guide forensic examiners to verify that the crime was committed with automated bots. Authors of Adepu, S., et al. [12] introduce a study to investigate the impact of cyber-attacks on a Water Distribution (WADI) system. The developed tool was able to launch multiple attacks and led to a better understanding of attack propagation and behavior of WADI in response to the attacks.

Moreover, most investigation research is based on access and system log files, but these logs are often massive, have a flat structure, contain very basic and limited information, and much of their data is not relevant to the investigation [11][13]. Nevertheless, a relatively old study from 2007 [13] describes an automated attack-tree-based algorithm for eliminating irrelevant information from system log files. The proposed approach also conducts systematic investigations of computer attacks.

Besides research proposals and tools, commercial solutions are also available. Gigamon GigaSECURE [14] enables security teams to gain broad access to and control over network data [15]. It uses five security systems to 1) aggregate network traffic, 2) detect the threat and generate events, 3) evaluate events, 4) generate related alerts, 5) capture the resulting process, and 6) investigate the results. On the other hand, Google has introduced a new security investigation tool, in 2018, as part of the Google workspace security center [16]. This tool aims to help admins and security analysts identify, classify, and take actions on their organization's threats.

Beyond that, there is also a lack of theoretical foundations and usage documentation to guide dashboard design, which a study conducted in 2019 tries to tackle. The "What Do We Talk About When We Talk About Dashboards?" report [17] performs a comprehensive analysis of dashboards by characterizing them and the practices surrounding them. This work is essential to support the correct design and implementation of different types of dashboards. Another study from 2018 describes a framework to analyze and develop dashboard templates for enterprises [18]. It focuses on the event/business operation, attribute list, visualization, dashboard capabilities, and the final evaluation of the dashboard.

## 3  Investigation Approach

Our investigation approach regards a specific security workflow and domain. Cyber sensors as well as physical sensors distributed across the system's infrastructure, collect, daily, vast amounts of events about system activity. To detect malicious behavior and probable attack attempts, event data must be first carefully processed in an efficient manner because it is impossible for a human to analyze such amount of data in an effective way. So, in order to solve this issue, the events originated from the multiple heterogeneous sources of the system are passed to a component designated as Correlation Engine. The Correlation Engine is a pattern matching mechanism that contains expert written rules which are periodically reviewed and updated under a strict protocol. When a set of events trigger a given rule, an alert is originated. Alerts are then sent to an Incident Management Portal (IMP) which provides a user-friendly interface for the SOC Operator. In the IMP, the SOC Operator can visualize generated alerts and query for more related information such as event details. The SOC, with the available information classifies the alerts as either incidents or not triggering an appropriate security response to handle the abnormality. The described systems and their interactions are represented in the UML diagram of Figure 1.

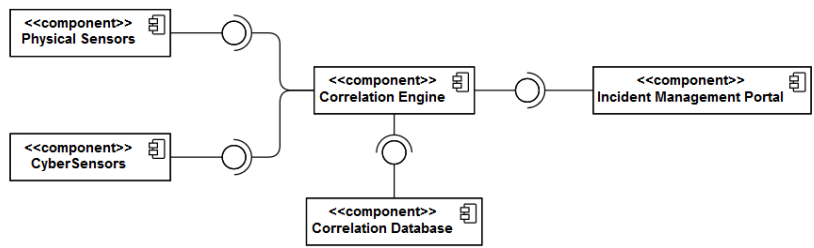

**Fig. 1.** Containers View – Security Ecosystem

The concerned security domain regards three key concepts:
**Event.** Contains low-level information about system activity. Events can have multiple heterogeneous sources, which can be either physical, such as closed-circuit television and access control management systems, or cyber, such as intrusion detection systems and firewall data. Dealing with events from heterogeneous sources is not a straightforward process and requires a lot of investment in terms of data representation and parsing.
**Alert.** Is originated when a set of events trigger a specific security rule. Alerts allow the correlation between events and contain summary information of those same events. They also include a brief justification about how they were generated.
**Incident.** An Incident results of an alert classification by the SOC Operator.

### 3.1 Architecture

Our solution, designated as SMS-I (Security Management Solutions – Investigation) is an investigation tool that acts as an enhanced, fully-automated decision support mechanism. SMS-I gathers information from both alerts and events, producing incident probabilities based on Machine Learning models. It also provides a set of functionalities to the SOC Operator so that he can perform an in-depth inspection about alert occurrences, such as timeline-based dashboards, alert listing, and incident occurrence probability filters. Hence, the SOC Operator when presented with an alert can visualize the probability of it being an incident based on the Machine Learning predictions and can easily perform a thorough inspection of its occurrence, facilitating his work. The SMS-I architecture is described in Figure 2.

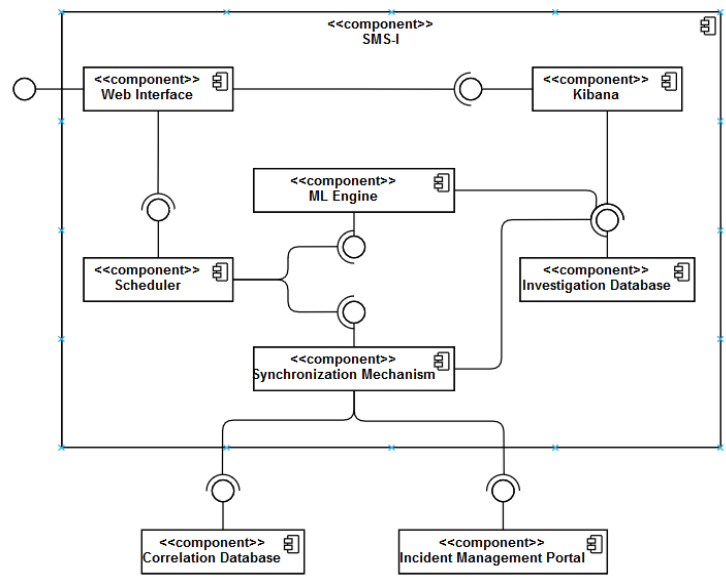

**Fig. 2.** Containers View – SMS-I Internal Architecture

The SMS-I comprises several internal modules:
**Synchronization Mechanism**. It is responsible for the data synchronization process, obtaining events and alerts from the Correlation Engine database and incidents from the Incident Management Portal, parsing them into predefined formats, and storing them in the Investigation Database.
**ML Engine**. The ML Engine is responsible for executing Machine Learning models to determine the probability of each Alert being an Incident based both on its own features and on the features of its related events. This information is also stored in the Investigation Database.
**Scheduler**. The Scheduler orchestrates both Synchronization Mechanism and ML Engine triggering their execution given some constraint (E.g., time-based - every minute).

**Investigation Database**. It contains the data of our system: events, alerts, incidents and Machine Learning results.

**Kibana**. Is the interface of the Investigation Database, it provides several visualization methods that we combined in order to produce intuitive and informative dashboards.

**Web Interface:** It encapsulates the Kibana Dashboards and provides additional functionalities such as interactive listings and filters.

### 3.2 Machine Learning

The ML Engine is a very important component of our architecture since it has the responsibility to execute the Machine Learning models and to store the produced results in the Investigation Database so that they can be accessed by both Web Interface and Kibana. In our work, Machine Learning is used from a supervised perspective and it can be detailed into two different processes, Learning and Predicting.

**Learning**. In the learning phase, events, alerts, and incidents are collected offline and preprocessed into a clean dataset in tabular format. The alerts contain a set of features such as severity, detector and related asset which are enriched with additional features engineered from its related events. The data from the IMP is used to label the alerts accordingly in order to identify which originated incidents and which did not. The produced dataset is then transformed into a comprehensible format so it can be used by the algorithm and split into three sets (train, validation and test) for evaluation purposes. After the training process, the model can use the broad patterns acquired from the data in order to recognize possible incidents on unseen alerts. Finally, the trained model is deployed into the ML Engine component.

**Predicting**. When the ML Engine is executed, it fetches the alerts (and related event data) from the Investigation Database which were not yet inspected by the algorithm. The Machine Learning model computes an incident probability score for each alert which are later stored in the Investigation Database.

Several algorithms such as Random Forest, Multi-layer Perceptron and Long Short-Term Memory were experimented. For a more detailed description of the employed methods readers can be redirected to [22].

## 4 Visualization Approach

A dashboard is a collection of visual displays of information used to quickly monitor conditions and increase general understanding [19].

Visualization dashboards are applied by almost every organization, including airport security centers, financial institutions, and healthcare providers, due to their increasing importance in a data-driven environment. Their application leads to improved decision-making processes, operational efficiency, and data visibility [17]. While designing the dashboards, it is important to consider that 50% of SOC analysts receive less than 20 hours of training for year [20]. Thus, there was a conscious effort to keep the visualizations simple and inviting to help the analysts' performance and their understanding of the investigation tool.

### 4.1 Dashboards

Given their diversity, dashboards can have different purposes, audiences, and visual elements. In this section, the visualization dashboards developed to aid cyber-physical security processes will be presented and justified.

**Technology Used**. Firstly, Elasticsearch is responsible for the analysis, normalization, enrichment and storage of alert and incident data, as well as data provided by machine learning algorithms. All this information is then accessed by Kibana to create new dashboards, which allows the user to search and visualize security data. Both tools belong to Elastic Stack, an open source framework and are easily integrated with one another. The decision to implement these tools is based on their flexibility and ease of installation, configuration, and execution.

Kibana represents the graphical interface of Elastic Stack. It applies and combines different visualization methods like graphics, tables, charts, and metrics to build dashboards. Furthermore, the Timelion plugin provided by Kibana allows the analysis of security data in a continuous timeline.

**Alerts Dashboard**. The alerts dashboard includes all data related to security alerts generated by the different threat detection tools available in the SOC. One of this dashboard's main goals is to monitor the quantity, nature, and severity of alerts, considering their incident prediction probability, which is calculated by the machine learning algorithm. It is important to note that, in this solution, an incident can only have one corresponding alert. However, an alert can be an incident, or it can be irrelevant.

According to a report developed by CriticalStart in 2019, more than 70% of analysts feel overwhelmed with the number of alerts and incidents they need to investigate for a day [20]. In comparison, more than 50% of organizations receive over 10,000 alerts daily, which can lead to alert fatigue and neglect [21]. So, to maintain SOC efficiency and reduce the impact of the investigation on the responsible personnel, it is essential to control the number of received alerts. The objective of the graphics and metrics shown in Figure 1 is to help avoid a sudden overload of alerts by monitoring the total number of cyber and physical alerts. An alert gauge was also developed to ensure that an overwhelming quantity of alerts is not reached.

*Alert Quantity Monitoring*. Figure 3 shows the first set of alert visualizations for the last sixty days.

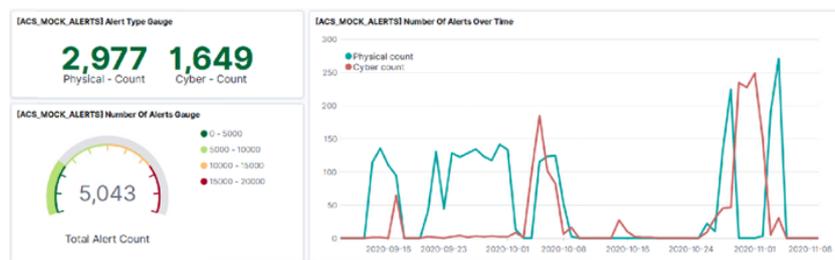

**Fig. 3.** Alerts Dashboard – Alert Quantity Monitoring Visualizations

The severity of an alert defines if it should be ignored or if there is a need to carry out a more thorough investigation. Four severity levels were defined, including high, medium, low, and info. Besides controlling the number of alerts for each severity level, to avoid the overburdening of SOC analysts, it is also possible to monitor the date of occurrence of alerts. This is useful to perform pattern and trend identification and to study previous incidents and preceding alerts.

*Alert Severity Monitoring.* Figure 4 displays the second important set of visualizations, which monitor the severity of alerts.

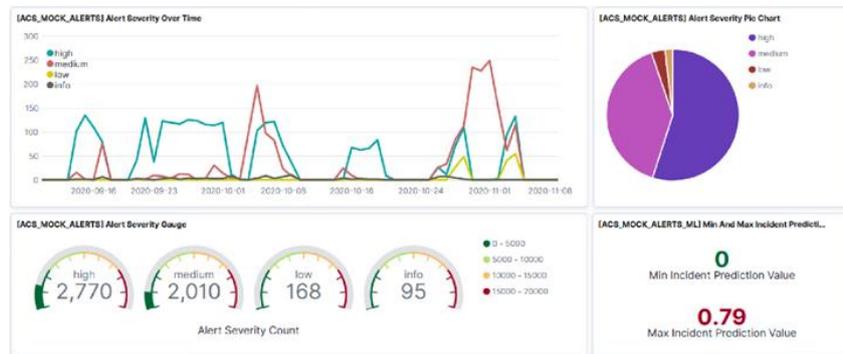

**Fig. 4.** Alerts Dashboard – Alert Severity Monitoring Visualizations

The set of graphics and metrics that incorporate Figure 5 display, from 0% to 100%, the number of alerts that possess a certain probability of being an incident, as well as the average incident prediction probability. In the example shown, most alerts have an incident prediction probability lower than 35%, which leads to a low average probability value. This means that overall, there probably is not any occurrence of an incident.

*Machine Learning Results.* Figure 5 shows the results provided by the machine learning algorithm. This includes the incident prediction probability, in other words, the likelihood of an alert representing an incident.

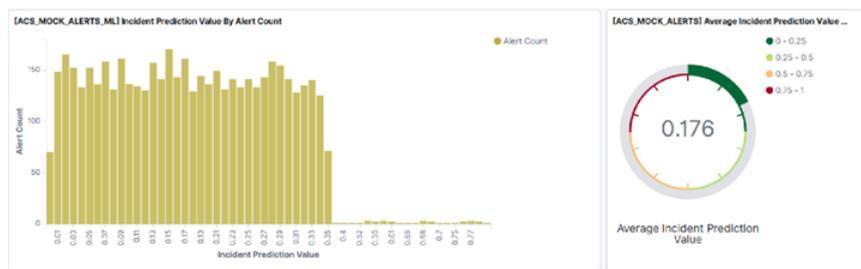

**Fig. 5.** Alerts Dashboard – Incident Prediction Probability Visualizations

**Incidents Dashboard.** The incidents dashboard aggregates all detected incidents related to the organization's security. This dashboard follows the structure of the alerts dashboard by monitoring the quantity, nature, and severity of incidents. However, since there were no incidents detected by the incident handling tools implemented in the SOC, there are no results to display on the dashboard.

### 4.2 Web Interface

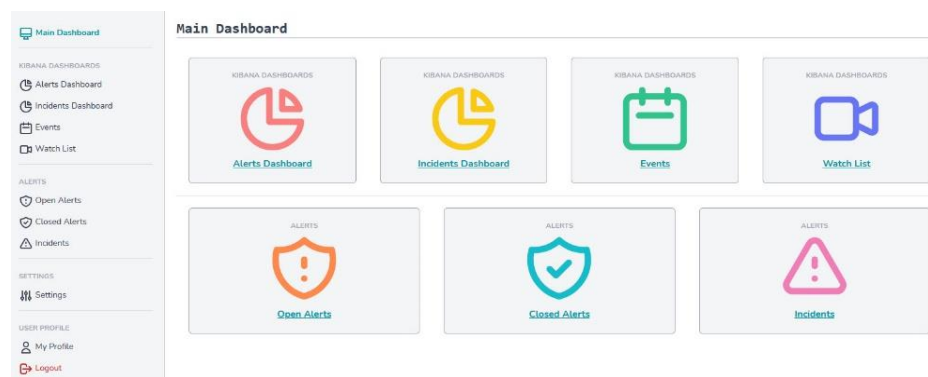

**Fig. 6.** Web Portal – Main Page

To facilitate the security investigation process, we developed a web interface for the investigation tool to extend monitoring, visualization, and evaluation of events taking place within the system. The investigation tool web interface is a dedicated web application that aims to grab all gathered information regarding security concerns and present them through a unified portal. Figure 6 represents the web portal's main page that contains intuitive links to its available high-level functionalities. Using this web app, security operations users can explore dashboards and information of different levels of details, including logical information, technical specifications, and machine learning forecasting results.

**Implementation and Setup.** The investigation tool web interface is a standalone web application built using MVC design pattern. It communicates with other internal modules using Restful APIs. The tool is a part of an internal secure network and provides limited access to a specific type of users, namely SOC operators, being authenticated using another sub-system.

**Main Features.** Following are the main features of the Investigation web application:

- Authentication and authorization of logged-in users using a remote authentication module;
- Access to interactive dashboards provided by Kibana module, these dashboards are embedded within the investigation tool and include alerts dashboard and incidents dashboard;
- Access to a list of alerts of the highest probability to be incidents, provided by the Investigation Database module;
- Access to lists of open alerts, closed alerts, and alerts categorized as incidents, provided by the Investigation Database module;

- Access to alert/incident specifications, including forecasting results retrieved from the Machine Learning Engine that are also stored in the Investigation Database;
- Access and control of user settings and manual trigger of the synchronization mechanism between other modules.

## 5    Conclusions

With this work, we describe how we are using machine learning and the ELK Stack to contribute to the investigation of cyber and physical attacks. A Visualization tool based on Kibana and a web portal, were developed to support Security Operators dealing with the amounts of information coming from both cyber and physical security monitoring solutions. Moreover, we also introduce to the SOC operator machine learning suggestions in order to help him to detect possible incidents. The ongoing work is now the application of this tool to different specific attack scenarios.